# Mid-infrared (3-8 μm) $Ge_{1-y}Sn_y$ alloys (0.15 < $y$ < 0.30): synthesis, structural, and optical properties

Chi Xu[1*], Patrick M. Wallace[1], Dhruve A. Ringwala[1], Shery L. Y. Chang[1,2], Christian D. Poweleit[3], John Kouvetakis[1], and José Menéndez[3]

[1]School of Molecular Sciences, Arizona State University, Tempe, AZ 85287-1604, USA

[2]Eyring Materials Center, Arizona State University, Tempe, AZ 85287-1704, USA

[3]Department of Physics, Arizona State University, Tempe, AZ 85287-1504, USA

$Ge_{1-y}Sn_y$ alloys with compositions in the 0.15 < $y$ < 0.30 range have been grown directly on Si substrates using a chemical vapor deposition approach that allows for growth temperatures as high as 290 °C. The films show structural properties that are consistent with results from earlier materials with much lower Sn concentrations. These include the lattice parameter and the Ge-Ge Raman frequency, which are found to depend linearly on composition. The simplicity of the structures, directly grown on Si, makes it possible to carry out detailed optical studies. Sharp absorption edges are found, reaching 8 μm near $y$ =0.3. The compositional dependence of edge energies shows a cubic deviation from the standard quadratic alloy expression. The cubic term may dramatically impact the ability of the alloys to cover the long-wavelength (8-12 μm) mid-IR atmospheric window.

[*]chixu@asu.edu

The recent development of Si-compatible $Ge_{1-y}Sn_y$ alloys represents an intriguing opportunity for infrared technologies, since the alloys are expected to possess a direct band gap $E_0$ between 0.8 eV and -0.4 eV, similar to the ubiquitous HgCdTe system. Furthermore, most theoretical predictions and extrapolations from experimental data indicate that $E_0$ becomes zero for $y = 0.25$-$0.30$, so that the two important mid-IR atmospheric windows in the 3-5 μm and 8-12 μm ranges should be accessible using Ge-rich alloys. However, a full experimental verification of these predictions is not available, because systematic band gap studies become increasingly problematic as the Sn concentration exceeds 15%. This is due to the fact that the standard Ge-buffer technology loses its effectiveness for accommodating the lattice mismatch, causing a deterioration in materials quality that can even lead to epitaxial breakdown.[1] Attempts to circumvent the quality issues to achieve alloys with $y \gg 0.15$ are based on lowering the growth temperature to about 150 °C,[2-5] or using complex buffer layers with intermediate compositions.[6-8] However, very few reports have been published on band gaps from such samples,[4,8,9] and the few results available are difficult to compare due to unknown or large strains present, compositional uncertainties, different methodologies for extracting the band gaps, and sample complexity. In this letter, we report on the structural and optical characterization of a series of $Ge_{1-y}Sn_y$ alloys that meet three criteria that reduce the band gap uncertainty and simplify the analysis of optical experiments: first, a smooth and monotonic compositional dependence of the structural properties that is consistent with previous measurements of low-Sn alloys with proven quality; second, small levels of strain that minimize errors associated with deformation potentials and elastic parameters[10] (and their unknown compositional dependence); and third, elimination of buffer layers that make it difficult to extract the optical properties of the $Ge_{1-y}Sn_y$ layer of interest.

Our $Ge_{1-y}Sn_y$ alloys were synthesized by CVD using stoichiometric reactions of high-reactivity $Ge_3H_8$ and $SnD_4$ custom reagents.[11] The layers are grown directly on Si, bypassing Ge buffers and/or complex graded layers. The composition range reaches far beyond the previous $y = 0.17$-$0.18$ threshold for samples grown directly on Si,[12] and includes the highest Sn levels synthesized to date using practical CVD methods. The growth is conducted between 245-290 °C, significantly above the temperatures (~150 °C) employed in MBE. This facilitates nearly full strain relaxation, as shown by X-ray diffraction (XRD). The relaxed lattice parameter and the Ge-Ge Raman frequency follow the same linear compositional dependence previously established in low-Sn films,[13,14] demonstrating similar structural properties and no Sn-segregation. The simplicity of the structures makes it possible to carry out detailed optical studies using spectroscopic ellipsometry (SE). In the visible range, we find sharp features corresponding to all optical transitions observed in Ge-like materials. In the IR, we find absorption edges extending all the way down to 8 μm. The compositional dependence of these features show a nearly ideal quadratic dependence for high-energy features, but clear deviations from this dependence for $E_0(y)$.

For a typical growth experiment, stock mixtures are prepared using 0.25 g of $Ge_3H_8$ and varying amounts of $SnD_4$ to achieve stoichiometric Sn/Ge fractions matching the desired alloy composition. We also explored the effect of doping levels of Si by adding small amounts of $Si_4H_{10}$ for some of the experiments. The mixture is diluted with 1.3 liters of $H_2$ and placed on the gas flow manifold of the deposition system. The substrates are RCA-cleaned quadrants of 100-mm Si(100) wafers. These are subsequently dipped in a 5% HF/MeOH bath, dried under a stream of $N_2$, and loaded into a quartz boat capable of accommodating multiple wafer segments positioned upright 1 cm apart. The boat is then placed into the load-lock of the deposition system, pumped down to $10^{-8}$ Torr and then inserted under a continuous flow of $H_2$ into the CVD chamber. The latter is a ⌀ 3" quartz tube externally heated by a three-zone resistance furnace. In preparation for growth the tube and boat are subjected to a Si coating to passivate the walls. The growth temperature is then set to the desired level and a background $H_2$ flow at 200 mTorr is established under dynamic pumping using oil-free pumps. This $H_2$ source serves as an additional diluent of the reaction mixture. After a short $Ge_2H_6$ clean, the reaction mixture is injected through calibrated mass flow controllers and combined with the $H_2$ carrier gas to commence growth. Typical deposition experiments lasted 60 min.

Rutherford backscattering (RBS) measurements following the methodology of Ref. 13 gave Sn/Ge ratios closely matching the corresponding ratios in the gaseous mixtures. In the Si-doped samples, the amount of Si was found to range from 2% to less than 1% at the highest Sn-concentrations. At such low levels Si has a very minor impact on the material properties,[15] but the ability to incorporate this element under the high-Sn growth conditions may turn out to be important to achieve full mid-IR coverage, as discussed below. The temperatures that maximize the growth rate while maintaining a mirror-like surface appearance and suppressing Sn precipitation was found to decrease monotonically from 285 °C ($y = 0.15$) to 245 °C ($y = 0.30$), while the corresponding film thicknesses ranged from 245 nm to at least 40 nm, respectively. For all of these samples, atomic force microscopy revealed uniformly planar surfaces devoid of dislocation-related features and rms roughness in the 3-5 nm range. We have not yet discovered an upper limit to the amount of Sn that can be incorporated using our growth method. Preliminary results show Sn concentrations as high as $y = 0.32$.

Figure 1 shows the XRD (004) and (224) maps for a $Ge_{0.75}Sn_{0.25}$ film grown on Si(100). The contours reveal strong and well-defined peaks characteristic of a material with diamond cubic structure oriented along the growth direction. The strain ε for "as grown" samples computed from these measurements was found to be slightly compressive (ε=-0.5% for $y = 0.13$) and decreased in magnitude for higher Sn concentrations, becoming marginally tensile (ε < 0.2%) or vanishingly small for $y > 0.17$. An example is seen in Fig. 1, where the relaxation line passes through the center of the (224) map, indicating that the layer is fully relaxed within error. Figure 2 shows the relaxed lattice constant $a_0$ versus Sn concentrations measured by RBS. The points agree well with the solid black line representing a linear interpolation between pure Ge and α-Sn (Vegard's law). The seamless consistency of the new data with the previous trend provides further evidence for the formation of single-phase alloys incorporating substitutional Sn. Similarly, the Raman scattering data in Figure 3 show a linear compositional dependence of the Ge-Ge Raman frequency with a slope consistent with previous measurements at lower Sn-concentrations.[14] The increasing asymmetric broadening as a function of $y$ is expected as a result of mass and bond disorder and does not necessarily imply the presence of crystalline defects.[16]

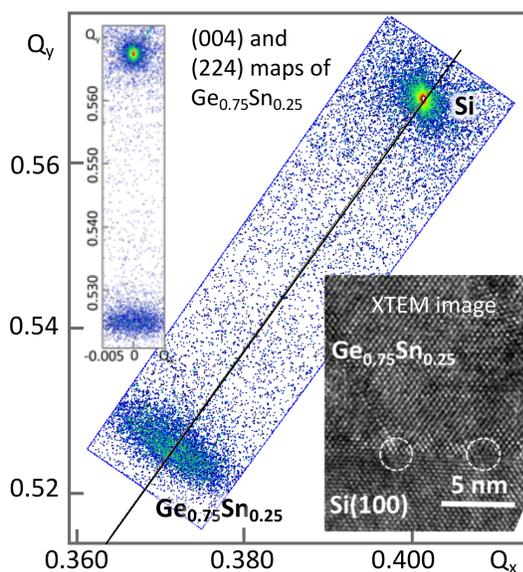

**Figure 1:** XRD (004) and (224) maps for a $Ge_{0.75}Sn_{0.25}$ film indicating good crystallinity. The (224) reciprocal space map reveals full relaxation of the misfit strain as evidenced by the relaxation line passing through the peak maximum. The inset shows a high resolution XTEM image and the location of edge dislocations marked by circles. The separation between the defects is consistent with the 8% lattice mismatch between film and substrate.

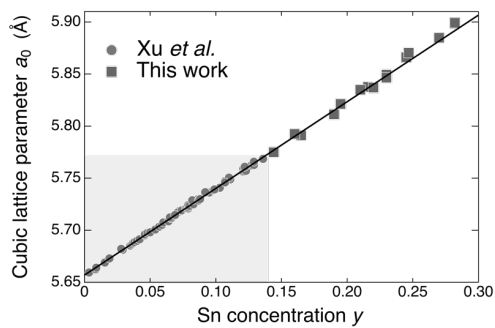

**Figure 2**: Compositional dependence of the relaxed lattice constant $a_0$ for $Ge_{1-y}Sn_y$ samples with $0 < y < 0.29$. The circular dots are previous results from Ref 13, and the black dots with a grey background correspond to current samples. The solid line denotes Vegard's law between the Ge and α-Sn end points.

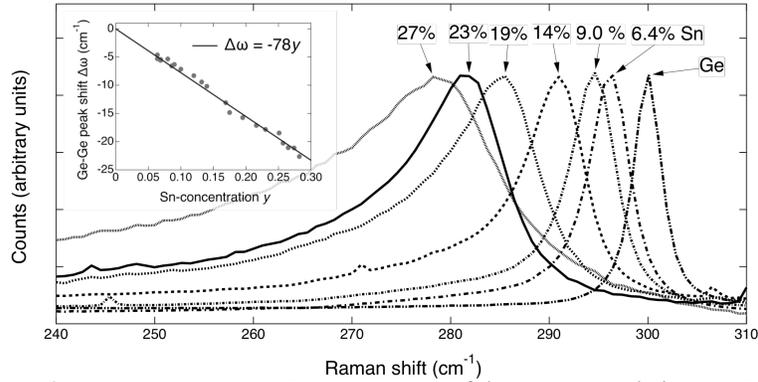

The film morphology and microstructure were further measured using cross-section transmission electron microscopy (XTEM). The experiments

**Figure 3** Room temperature Raman spectra of the Ge-Ge mode in $Ge_{1-y}Sn_y$ alloys obtained with 514 nm excitation in the $z(x,y)\bar{z}$ scattering configuration, where $x$, $y$ and $z$ correspond to the cubic cartesian axes. The peak maximum has been normalized for display purposes. For this experiment we grew additional samples with $y < 0.15$. The inset shows the compositional dependence of the Raman shift, in excellent agreement with previous results for low Sn-concentration as compiled by Gassenq *et al*. (Ref. 14)

show that strain relaxation occurs mainly by formation of edge dislocations confined to the interface plane. This is shown in the Fig. 1 inset, which features a high-resolution view of the $Ge_{0.75}Sn_{0.25}$/Si interface region in (110) projection. Defects are marked by circles. The image also shows the expected epitaxial commensuration of the (111) lattice fringes, corroborating the good crystal alignment observed by RBS and XRD. Figure 4 shows a bright field image of a $Ge_{0.78}Sn_{0.22}$ film illustrating a homogeneous contrast throughout the alloy epilayer, indicative of single-phase material. The layer is thick (over 100 nm) and uniform with a reasonably flat surface and a sharp well-defined transition at the interface. The latter is epitaxial as shown in the high-resolution image inset.

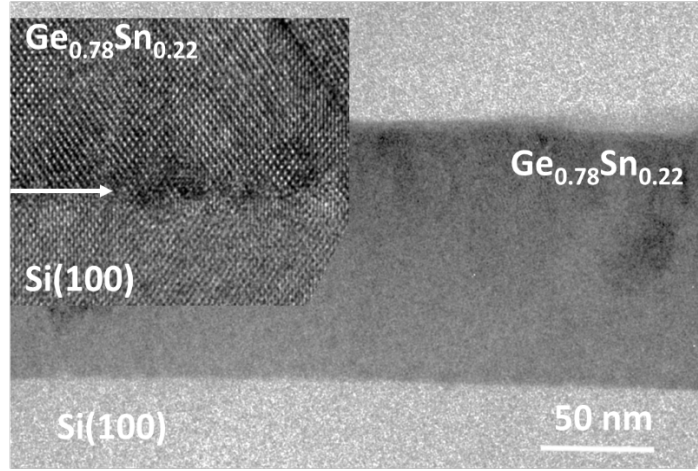

**Figure 4**: XTEM data of a $Ge_{0.78}Sn_{0.22}$ showing a bright field image of the full layer and a high resolution cross sectional view of the heterointerface marked by the arrow in the inset. The layer thickness is 100 nm.

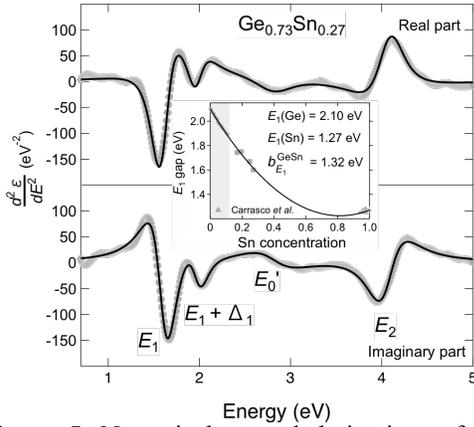

**Figure 5**: Numerical second derivatives of the NIR-visible-UV dielectric function of a $Ge_{0.73}Sn_{0.27}$ film (grey circles) and theoretical fit including four interband transitions as indicated. The inset shows the compositional dependence of the $E_1$ transition. The small grey circles over a grey background are from Ref. 17, the grey squares are from this work, and the grey triangles are from Ref. 19. The solid line is the best quadratic fit from Ref. 17, which also agrees well with the extended data presented here.

Spectroscopic studies of above band-gap optical transitions were carried out at room temperature using a J.A. Woollam™ UV-Vis variable angle spectroscopic ellipsometer. The ellipsometric data collection and processing was carried out as described in Ref. 17. Figure 5 shows an example of numerical second derivatives of the real and imaginary part of the dielectric function, which were obtained using a combination of a regularization method and a Savitzky-Golay filter as described in Ref. 18. Sharp transitions are observed, and an excellent fit of the data can be obtained using standard expressions for Ge-like materials in this energy range. This makes it possible to identify all expected transitions, as indicated in the figure. The ability to unambiguously identify these features corroborates the high crystallinity of the alloys.

The compositional dependence of optical transition energies in alloy semiconductors such as $Ge_{1-y}Sn_y$ deviates from a linear interpolation between the Ge and α-Sn values. This deviation is usually well fit by a quadratic term of the form $-by(1-y)$, where $b$ is the so-called bowing parameter. The inset shows the case of the $E_1$ transition. In Ref. 17 a value $b_1 = 1.32$ eV was obtained for the bowing parameter of this transition by fitting experimental data from $Ge_{1-y}Sn_y$ alloys with $y < 0.14$. These data points appear in the shaded region of the inset. The data from our new samples with $0.14 < y < 0.30$, shown by squares, fall almost perfectly on the same curve determined from the low Sn-concentration data. Essentially the same bowing parameter was also found in Sn-rich samples by Carrasco *et al.*, (Ref. 19) so that the quadratic dependence is valid over the entire $0 < y < 1$ compositional range.

The band gap spectral region was investigated with infrared SE (IRSE). The measurements were performed on a J.A. Woollam™ IR-VASE system over an energy range extending from 0.03-0.7 eV, with a step size of 1meV and three angles of incidence, typically 65°, 70°, and 75°. The sample was modeled as substrate, a GeSn film, an oxide layer, and a roughness layer. The ellipsometric thicknesses were consistent with the RBS and XTEM measurements. A two-step fit of the data, as carried out for the visible range, allowed the extraction of the real and imaginary part of the dielectric function without resorting to any *a priori* theoretical model. From the dielectric function we computed the absorption coefficient, which we show in Fig. 6(a) for selected samples. The data show clear absorption edges that shift monotonically to lower energies as the Sn concentration is increased, reaching 8 μm for $y \sim 0.3$. To confirm that the observed edge corresponds to interband transitions at the direct gap $E_0$, we computed the absorption coefficient using a method similar to the one introduced in Refs. 20 and 21. We show the result for the highest-Sn concentration sample as a dotted line in Fig. 6(a). The good agreement between theory and experiment, both for pure Ge (Refs. 20 and 21) and Ge$_{1-y}$Sn$_y$ provides strong support for our interpretation in terms of interband transitions. It is important to emphasize that our theoretical expressions contain the value of $E_0$ and its broadening as adjustable parameters, but do *not* include an "amplitude" parameter that adjusts the absorption strength, as is often the case in the literature. Instead, all prefactors are computed for pure Ge using the experimental effective masses in this material, and extrapolated to Ge$_{1-y}$Sn$_y$ using k·p theory and assuming that the momentum matrix element $P$ scales as $a_0^{-1}$ (Ref. 22). For narrow band gap semiconductors at room temperature several additional effects must be accounted for that play a lesser role in pure Ge. These include screening of the excitonic interaction by thermally activated carriers, Burstein-Moss shifts, and band structure non-parabolicity. We have accounted for these effects by adapting the analytical expression for the complex dielectric function obtained by Tanguy for the so-called Hulthen excitonic potential.[23] This potential includes a screening parameter $g$ that is computed using a prescription from Bányay and Koch[24] starting from the Thomas-Fermi screening wave vector, which is calculated using standard expressions.

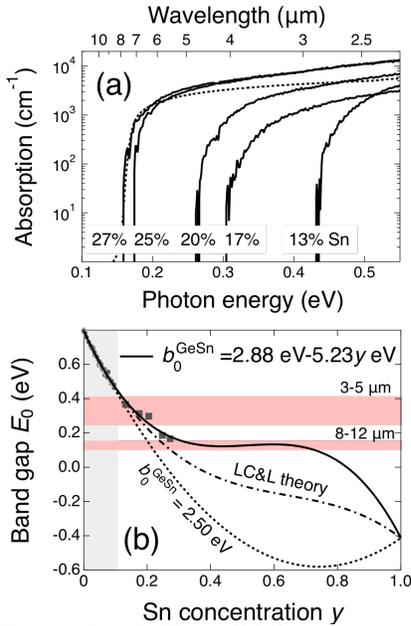

**Figure 6:** (a) The solid lines represent the experimental absorption coefficient from selected Ge$_{1-y}$Sn$_y$ alloys. The compositions are indicated at the bottom of the traces. (b) Compositional dependence of the direct gap $E_0$. The small grey circles are from Ref. 25, and the grey squares correspond to this work. The solid line is a global cubic fit of the data, the dotted line is a quadratic fit from Ref. 25, and the dashed-dotted line is the prediction from Ref. 28. The red horizontal bands highlight the two mid-IR atmospheric windows.

The relaxed direct band gap $E_0$ as a function of composition is shown in Fig. 6(b). Most experimental measurements and theoretical predictions of $E_0$ ($y$) are well represented by quadratic expressions containing a bowing parameter $b_0$. However, Gallagher et al found that the bowing parameter is itself compositionally dependent,[25] and proposed an expression of the form $b_0 = b_0^{(0)} + b_0^{(1)} y$, with $b_0^{(0)} = 2.66$ eV, and $b_0^{(1)} = -5.4$ eV. These fit parameters were obtained by studying samples with $y < 0.10$, which are indicated as circles in Fig. 6(b). Convincing statistical evidence for this effectively cubic compositional dependence was obtained from a very large sample set, but the deviations from a purely quadratic function are very small for $y < 0.10$. Our extended data for $y > 0.10$, on the other hand, provide clear evidence for the characteristic S-shape associated with cubic terms. A fit that includes all available data points gives $b_0^{(0)} = 2.88 \pm 0.04$ eV, and $b_0^{(1)} = -5.23 \pm 0.025$ eV, and is shown as a solid line. The S-like shape in the compositional dependence has been observed in III-V alloy systems[26,27] and was justified theoretically in Ref. 25. Furthermore, Lan, Chang, and Liu (LC&L) also predict an S-like dependence from a pseudopotential band structure calculation within the virtual crystal approximation.[28] Their calculation is perfectly fit using a cubic expression, and we show the result (combined with room temperature band gaps) as a dash-dotted line in Fig. 6(b). We see that the deviations from the best fit to the data and the LC&L prediction are not very large for the $y < 0.3$ available experimental data, but the extrapolation to Sn-rich alloys are significantly different. Both predictions seem to disagree with recent measurements on samples with $y > 0.94$ (Ref. 19), but the discrepancy should be interpreted with caution because the $E_0$ transition in α-Sn has a unique line shape that is not fully understood.[29]

The difference between the LC&L prediction and the best fit in Fig. 6(b) has an important practical consequence: LC&L predict a vanishing band gap for $y = 0.35$, which implies that the 8-12 μm window could be easily covered with Ge-rich Ge$_{1-y}$Sn$_y$ alloys. On the other hand, if our best fit is valid well beyond $y > 0.3$, the full 8-12 μm window would only be accessible to alloys with $y = 0.7$-0.8, which, if feasible, may require a completely different growth strategy. There is, however, a counterintuitive approach that may lead to smaller band gaps in a Ge-rich material: the incorporation of Si. The zone-center direct band gap of Si is $E_0 = 4.1$ eV.(Ref. 30) Therefore, adding Si should in principle raise $E_0$ dramatically. However, the compositional dependence $E_0$ ($x,y$) of the ternary alloy (where $x$ is the Si concentration) includes a term $-b_0^{SiSn}xy$ (Ref. 31), so that if $b_0^{\text{SiSn}} y > E_0^{\text{Si}} - E_0^{\text{Ge}}$, the addition of Si *lowers* the band gap. Experimental values of $b_0^{\text{SiSn}}$ range from 13 eV to 28 eV.(Refs. 32,33,34) so that for $y > 0.3$ the addition of Si should indeed reduce $E_0$. Our finding that Si can be incorporated into the lattice using the $Si_4H_{10}$ precursor suggest that our growth strategy is a promising route to achieve full mid-IR coverage at modest Sn concentrations.

This work was supported by the AFOSR under grants FA9550-17-1-0314 and FA8650-18-C-1152. The use of the TEM facility at the Eyring Materials Center is gratefully acknowledged. We thank Prof. Stefan Zollner for helpful discussions and data sharing.